\documentclass[aip,amsmath,amssymb,reprint,numerical,graphicx]{revtex4-1}
\usepackage{graphicx}% Include figure files
\usepackage{dcolumn}% Align table columns on decimal point
\usepackage{bm}% bold math
\begin{document}
\title{Shapiro like steps reveals molecular nanomagnets' spin dynamics}
\author{Babak Abdollahipour}
\email{b-abdollahi@tabrizu.ac.ir}
\affiliation{Faculty of Physics, University of Tabriz, Tabriz
51666-16471, Iran}
\author{Jahanfar Abouie}
\email{jahan@iasbs.ac.ir}
\affiliation{Department of Physics, Institute for Advanced Studies
in Basic Sciences (IASBS), Zanjan 45137-66731, Iran}
\author{Navid Ebrahimi }
\affiliation{Department of Physics, Institute for Advanced Studies
in Basic Sciences (IASBS), Zanjan 45137-66731, Iran}
\begin{abstract}
We present an accurate way to detect spin dynamics of a nutating
molecular nanomagnet by inserting it in a tunnel Josephson junction
and studying the current voltage (I-V) characteristic. The spin
nutation of the molecular nanomagnet is generated by applying two
circularly polarized magnetic fields. We demonstrate that modulation
of the Josephson current by the nutation of the molecular
nanomagnet's spin appears as a stepwise structure like Shapiro steps
in the I-V characteristic of the junction. Width and heights of
these Shapiro-like steps are determined by two parameters of the
spin nutation, frequency and amplitude of the nutation, which are
simply tuned  by the applied magnetic fields.
\end{abstract}
\maketitle
Molecular nanomagnets have recently attracted intense attentions due
to their large effective spin and long magnetization relaxation
times\cite{Gatt07} which make them suitable for applications in
quantum information processing\cite{Leuenberger01} and molecular
spintronics.\cite{Bogani08,Rocha05} A crucial aspect of the research
on molecular nanomagnets is determination of their spin dynamics.
One of the possible methods to detect the spin dynamics of molecular
nanomagnets is ferromagnetic-resonance experiment which is
extensively used for thin ferromagnetic layers.\cite{Bell08} Another
powerful technique to investigate the spin dynamics in bulk samples
is inelastic neutron scattering. In molecular nanomagnets, the
dynamics are usually extrapolated by fitting inelastic neutron
scattering spectra to a spin Hamiltonian and by performing
calculations within the framework set by this model.\cite{Baker12}
Transport measurement through nanomagnets is an on-demand method to
determine the magnetic state of a nanomagnet.\cite{Burzuri12} An
intrinsic limitation to the current measurement is associated to the
high access resistance of the normal contacts. However, when the
contacting leads become superconducting, long-range correlations can
extend throughout the whole system by means of the proximity effect.
This not only lifts the resistive limitation of normal-state
contacts, but further paves the way to probe electron transport
through a single
molecule.\cite{Lee08,Sadovskyy11,Kasumov05,Winkelmann09,Gaudenzi15}

Mutual interaction of the Josephson current flowing through the
junctions and the molecular nanomagnet's spin dynamics elicits
several interesting phenomena such as the modulation of the
Josephson current\cite{Zhu03} and generation of a circularly
polarized {\it ac} spin current with the nanomagnet's Larmor
precession frequency.\cite{Teber10} The effects of molecular
nanomagnet's spin nutation on the Josephson current flowing through
the Josephson junction has been studied in Ref.\cite{Abouie13} It
has been shown that the spin nutation of the molecular nanomagnet
causes generation of an $ac$ Josephson current through the junction,
in addition to the $dc$ one.
%
%%%%%%%%%%%%%%%%%%%%%%%%%%%%%%%%%%%%%%%%%%%%%%%%%%%%%%%%%%%%%%%%%%%%
%%%%%%%%%%%%%%%%%%%%%      Fig 1     %%%%%%%%%%%%%%%%%%%%%%%%%%%%%%%
\begin{figure}
\centerline{\includegraphics[width=7cm]{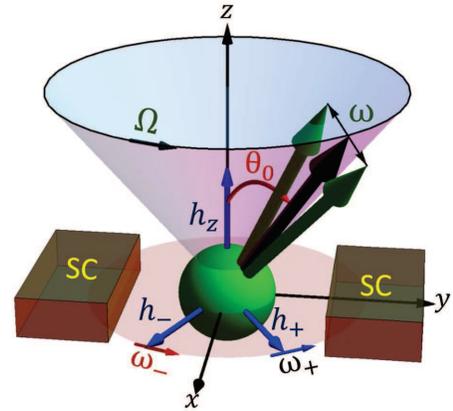}} \caption{(Color
online) A Josephson tunnel junction with a molecular nanomagnet. Two
circularly polarized applied magnetic fields (blue arrows) result in
the nutation of the molecular nanomagnet's spin (green arrows) which
has been schematically shown in the figure. $h_z$ controls the
precession frequency $\Omega$, and the nutation frequency $\omega$
is determined by frequencies of the $h_+$ and $h_-$.}
\label{system-model}
\end{figure}
%%%%%%%%%%%%%%%%%%%%%%%%%%%%%%%%%%%%%%%%%%%%%%%%%%%%%%%%%%%%%%%%%%%%
%

Irradiation of a microwave to a Josephson junction eventuates in
{\it Shapiro step} structure \cite{Shapiro63} in the current-voltage
(I-V) characteristic. In thin ferromagnetic layer/superconductor
Josephson junctions Shapiro steps reveal the magnetic response of
the ferromagnet.\cite{Petkovic09,Volkov09,Hikino12} They appear at
voltages $V=n(\hbar/2e)\Omega$ with the ferromagnetic resonance
frequency $\Omega$, integer $n$, and the ratio of the Plank constant
and the elementary charge, $\hbar/e$. A small magnet in the
proximity of a weak link with purely electromagnetic mutual
interaction may be detected through Shapiro-like steps caused by the
precession of the magnetic moment. The magnetic field of the
nanomagnet alters the Josephson current flowing through the link,
while the magnetic flux generated by the Josephson junction acts on
the magnetic moment of the nanomagnet.\cite{Cai10} The interplay
between the $ac$ Josephson current and the magnetization precession
of the nanomagnet shows a spin-polarized Shapiro steps and rich
subgap structure in the I-V curve.\cite{Holmqvist14}

In this paper, we proffer a feasible way to detect directly the spin
dynamics of a nutating molecular nanomagnet by current measurement.
By inserting the molecular nanomagnet in a tunnel Josephson
junction, and investigating behavior of the current flowing through
the junction we demonstrate that at low temperatures and in the
tunneling limit the modulation of the Josephson current due to the
{\it nutation} of the molecular nanomagnet, obviously seen as a
stepwise structure in the I-V characteristic. We show that these
Shapiro-like steps disappear when the molecular nanomagnet has only
a precession around $z$-axis.

{\it Spin nutation}: Spin nutation of the molecular nanomagnet is a
result of its interaction with the following effective magnetic
field:
\begin{equation}
\mathbf{h}_{eff}(t)=\mathbf{h}_{z}+\mathbf{h}_+(t)+\mathbf{h}_-(t)\ ,
\label{eff-field}
\end{equation}
where $\mathbf{h}_{z}$ is a static field along $z$ axis and includes
the external magnetic field as well as all other contributions such
as exchange interaction, crystal anisotropy, and magnetostatic
interaction, and $\mathbf{h}_{\pm}(t)
(=-\frac{h_{xy}}{2}[\cos(\omega_{\pm}t)\hat{x}+\sin(\omega_{\pm}t)\hat{y}])$
are two circularly polarized fields with amplitude
$\frac{h_{xy}}{2}$ and frequencies $\omega_{\pm}$. (see
Fig.(\ref{system-model})) Using the phenomenological Landau-Lifshitz
equation, the spin dynamics of the molecular nanomagnet, in
the absence of spin relaxation processes, is obtained as
\begin{equation}\label{spin-smm}
\mathbf{S}(t)=S\left(\sin\theta(t)\cos\Omega t,\sin\theta(t)\sin\Omega
t,\cos\theta(t)\right)\ ,
\end{equation}
where $S$ is magnitude of the molecular nanomagnet's spin, $\Omega$
$(=(\omega_+-\omega_-)/2=\gamma h_z)$ is the precession frequency
around $z$ axis with gyromagnetic ratio $\gamma$, and $\theta(t)$ is
the time-dependent tilt angle of ${\mathbf S}$ which is given by
$\theta(t)=\theta_0-\delta\theta\cos\omega t$. Here, $\delta\theta$
$(=\gamma h_{xy}/\omega)$ and $\omega$ $(=(\omega_++\omega_-)/2)$
denote the amplitude and frequency of the nutation of the molecular
nanomagnet's spin around $\theta_0$, respectively.

{\it Josephson current}: Hamiltonian of the tunnel Josephson
junction, schematically shown in Fig. (\ref{system-model}), is given
by:
\begin{equation}\label{total-hamiltonian}
H(t)=H_L+H_R+H_T(t)\ ,
\end{equation}
where $H_L$ and $H_R$, are the BCS Hamiltonian of the left (L) and
right (R) superconducting leads with identical amplitude of the pair
potential $\Delta$ and phases $\chi_L$ and $\chi_R$. These
Hamiltonians are written as:
\begin{equation}\label{hamiltonian-leads}
H_{\alpha}=\sum_{k,\sigma=\uparrow,\downarrow}\varepsilon_k
c_{\alpha k\sigma}^{\dagger}c_{\alpha
k\sigma}+\sum_k\left(\Delta_{\alpha} c_{\alpha
k\uparrow}^{\dagger}c_{\alpha -k\downarrow}^{\dagger}+h.c.\right)\ .
\end{equation}
where $\varepsilon_k$ is the energy of a single conduction electron, and
$c_{\alpha k\sigma}^{\dagger}(c_{\alpha k\sigma})$ is the creation
(annihilation) operator of an electron in the lead $\alpha=L, R$
with momentum $k$ and spin $\sigma$. The two superconducting leads are weakly
coupled via the tunneling Hamiltonian;
\begin{equation}\label{tunneling-hamiltonian}
H_T(t)=\sum_{k,k,\sigma, \sigma'}\left(c_{R k \sigma}^{\dagger}
T_{\sigma\sigma'}(t)c_{L k' \sigma'}+h.c.\right)\ ,
\end{equation}
where $T_{\sigma\sigma'}(t)$ is a component of the time dependent
tunneling matrix which transfers electrons through the system. The
tunneling matrix can be written as:
\begin{equation}\label{tunneling-matrix}
\hat{T}(t)=T_0\hat{\mathbf{1}}+T_S\hat{\mathbf{S}}(t)\cdot\mbox{\boldmath
$\sigma$}\ ,
\end{equation}
where $\hat{\mathbf{1}}$ is $2\times 2$ unit matrix,
$\hat{\mathbf{S}}(t)$ is the unit vector along the molecular
nanomagnet's spin (Eq. (\ref{spin-smm})), and $\mbox{\boldmath
$\sigma$}=(\sigma_x,\sigma_y,\sigma_z)$ denotes vector of the
Pauli's spin operators. The parameter $T_0$ is spin independent
transmission amplitude while $T_S$ denotes the amplitude of spin
dependent transmission originating from exchange interaction between
conduction electrons and the molecular nanomagnet's
spin.\cite{Teber10}

In the tunneling limit the Josephson current reads:
\begin{equation}\label{Josephson-current-operator}
I_{\alpha}^{\rm J}(t)=-e\int_{-\infty}^{t}dt'
\left(\left\langle\left[A_{\alpha}(t), A_{\alpha}(t')\right]\right\rangle+h.c.\right)\ ,
\end{equation}
where
\begin{equation}\label{a-operator}
A_{\alpha}(t)=\sum_{k,k',\sigma,\sigma'}c_{\alpha' k\sigma}^{\dagger}(t)
T_{\sigma\sigma'}(t)c_{\alpha k'\sigma'}(t)\ .
\end{equation}
In order to obtain an explicit form for $I_{\alpha}^{\rm J}(t)$ we define
the following retarded potential
\begin{equation}\label{X-definition}
X^{\sigma\sigma'}_{\rho\rho'}(t-t')=-i\Theta(t-t')\left\langle\left[a_{k,k'}^{\sigma\sigma'}(t)
,a_{p,p'}^{\rho\rho'}(t')\right]\right\rangle\ ,
\end{equation}
where $a_{k,k'}^{\sigma(\rho)\sigma'(\rho')}(t)=c_{\alpha'
k\sigma(\rho)}^{\dagger}(t) c_{\alpha k'\sigma'(\rho')}(t)$ and
$\sigma, \sigma', \rho$ and $\rho'$ denote $\uparrow$ and
$\downarrow$. This potential includes both triplet and singlet
correlations. In the presence of a spin-active junction including
spin-flip processes singlet correlations penetrating into the
magnetic region convert to triplet
correlations.\cite{Bergeret05,Braude07,Eschrig08} Moreover, the
interplay of the magnetization dynamics and transported carriers
also results in the conversion of the spin-singlet to the
spin-triplet correlations, which is accompanied by the absorption or
emission of a magnon.\cite{Takahashi07,Houzet08} These induced
triplet correlations have the same magnitude as the singlet
correlations at the interface, however they survive over a long
range despite the singlet correlations. The triplet correlations
also penetrate into the superconductors which have small amplitudes
in comparison to the bulk singlet components.\cite{Halterman07}

As the junction considered here has two tunnel barriers, we can
ignore the small effects of the triplet correlations induced in the
superconducting leads. Thus, the retarded potential
(\ref{X-definition}) simplifies to
$X^{\sigma\sigma'}_{\rho\rho'}(t-t')=\sigma\sigma'\delta_{\sigma,-\rho}
\delta_{\sigma',-\rho'}X_{ret}(t-t')$, where $\sigma,\sigma'=\pm 1$.
To the first order of the nutation amplitude ($\delta\theta$),
the Josephson current is obtained as:
\begin{eqnarray}
I_{\rm J}(t)=I_c\sin\chi=(I_0+I_1\delta\theta\cos\omega t)\sin\chi\ ,
\label{Josephson-current}
\end{eqnarray}
where $\chi$ $(=\chi_R-\chi_L)$ is the phase difference between two
superconducting leads. By defining $T_{||}=T_S\cos\theta_0$ and
$T_{\bot}=T_S\sin\theta_0$, the coefficients $I_0$ and $I_1$ are
given by;
\begin{eqnarray}
&&\nonumber I_0=2e\left[2(T_0^2-T_{||}^2){\cal R}(0)
-2T_{\bot}^2{\cal R}\left(\frac{\hbar\Omega}{2\Delta}\right)\right]\ ,\\
&&I_1=-2eT_{\bot}T_{||}\bigg[2{\cal R}(0) -2{\cal
R}\left(\frac{\hbar\Omega}{2\Delta}\right) +2{\cal
R}\left(\frac{\hbar\omega}{2\Delta}\right)\nonumber\\
&&-{\cal
R}\left(\frac{\hbar\omega+\hbar\Omega}{2\Delta}\right)-{\cal
R}\left(\frac{\hbar\omega-\hbar\Omega}{2\Delta}\right)\bigg]\ ,
\label{JC_Coefficients}
\end{eqnarray}
where ${\cal R}(x)$ $(=\Re\sum_{k,k'}X_{ret}(x))$ is the real part
of the retarded potential which is obtained using the Matsubara
Green's functions and analytical
continuation\cite{Mahan90,Teber10,Abouie13} as;
\begin{equation}\label{Realpart-X}
{\cal R}(x)=\left\{
\begin{array}{cl}
\pi N^2\Delta K(x)\;& x<1
\\
\pi N^2\frac{\Delta}{x}K(\frac{1}{x})\;& x>1
\end{array}\right.\ .
\end{equation}
Here, $N$ is density of states at the Fermi energy in the left and
right leads and $K(x)$ is the complete elliptic integral of the
first kind.
%%%%%%%%%%%%%%%%%%%%%%%%%%%%%%%%%%%%%%%%%%%%%%%%%%%%%%%%%%%%%%%%%%%%
%%%%%%%%%%%%%%%%%%%%%      Fig 2    %%%%%%%%%%%%%%%%%%%%%%%%%%%%%%%
\begin{figure}
\centerline{\includegraphics[width=8cm]{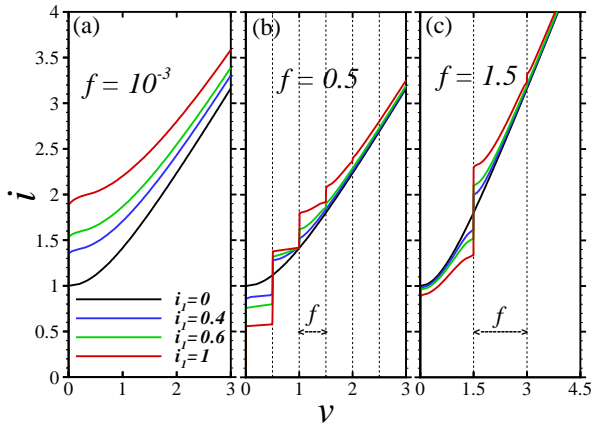}} \caption{(Color
online) I-V characteristic of the tunnel Josephson junction with a
nutating molecular nanomagnet. $f=\hbar\omega/2eRI_0$, $i_1=I/I_0$
and $v=V/RI_0$. Nutation of the nanomagnet leads to a stepwise
structure (Shapiro-like steps) on the I-V curve. Width of the steps
are equal to $f (\propto\omega)$ and their heights are proportional
to $i_1(\propto\delta\theta)$.} \label{I-V}
\end{figure}
%%%%%%%%%%%%%%%%%%%%%%%%%%%%%%%%%%%%%%%%%%%%%%%%%%%%%%%%%%%%%%%
%

In the absence of the circularly polarized magnetic fields,
$\delta\theta$ is zero and the spin of the molecular nanomagnet has
only a precession around $z$ axis. This eventuates in flowing of a
$dc$ Josephson current through the junction. Applying the two
circularly polarized magnetic fields $\mathbf{h}_{\pm}(t)$, leads to
the spin nutation of the molecular nanomagnet ($\delta\theta\neq0$),
and causes the generation of an $ac$ Josephson current through the
junction, in addition to the $dc$ one. We interpret this effect as
being due to evolution of Andreev bound states generated by the
correlated Andreev reflections from the superconducting interfaces.
Spin precession of the molecular nanomagnet affects the charge
current by making transitions between the continuum states below the
superconducting gap edge and the Andreev
levels.\cite{Holmqvist11,Holmqvist12} When the molecular
nanomagnet's spin has a nutational motion, the Andreev bound states
energies vary with the tilt angle oscillation which eventuate in the
modulation of the Josephson current.\cite{Abouie13}

{\it Phase dynamics}: In order to detect the spin dynamics of a
nutating molecular nanomagnet we insert it in a tunnel Josephson
junction and investigate the phase dynamics of the junction. The
phase dynamics of a tunnel Josephson junction composed of two
superconducting leads, having phase difference $\chi$, coupled
through a tunnel barrier and subjected to a bias current $I$ is
generally described by resistively shunted Josephson junction (RSJ)
model. In RSJ model, a dissipative Josephson junction is modeled by
a parallel circuit consisting of an ideal Josephson junction and a
resistance $R$.\cite{Petkovic09} The shunted resistance $R$ is
indeed used to model leakage currents through the junction. By
defining $I_c$ as the critical Josephson current, the phase dynamics
of the Josephson junction is governed by the following equation;
\begin{eqnarray}
I=I_c\sin\chi+\frac{\hbar}{2 e R}\frac{d\chi}{dt}\ ,
\label{RSJ_model}
\end{eqnarray}
where $I_c\sin\chi$ is Josephson current through an ideal Josephson
junction and the second term represents the current passing thought
the resistance $V/R$, where we have used the Josephson relation
$V=(\hbar/2e)d\chi/dt$. For a tunnel Josephson junction which has
not driven externally, the critical Josephson current $I_c$ depends
just on the junction parameters and it has no time dependence. For
such a junction and in the case that $I>I_c$, the phase difference,
given by $\chi=\arcsin(I/I_0)$, has not time dependence.
Consequently, the voltage difference is zero. For $I>I_c$ the phase
difference grows with time which results in a nonvanishing voltage
difference. In this case the current-voltage relation takes the form
$\langle V\rangle=R\sqrt{I^2-I_0^2}$ and it almost has a step at
$V=0$. If the junction is driven by an $ac$ current such that
$I=I_0+I_1\cos(\omega t)$, here $I_0$ is a static current and $I_1$
is the amplitude of the oscillation around it, the I-V
characteristic of the junction show an stepwise structure called
Shapiro step. These steps occur precisely when the average voltage
match $\langle V\rangle=n(\hbar\omega/2e$), where $n$ stands for
integer values.

Making use of the Josephson current given by Eq.
(\ref{Josephson-current}) the phase dynamics of the tunnel Josephson
junction driven by the external magnetic fields (Eq.
(\ref{eff-field})) is given by the following rescaled equation;
\begin{eqnarray}
i=(1+i_1\cos f t')\sin\chi+\frac{d\chi}{dt'}\ ,
\label{scaled-RSJ_model}
\end{eqnarray}
where $i (=I/I_0)$ is the dimensionless bias current,
$i_1=\delta\theta I_1/I_0$, $f=\omega/\alpha$, $t'=\alpha t$, and
$\alpha=2eRI_0/\hbar$. Solving Eq. (\ref{scaled-RSJ_model})
numerically the rescaled voltages $v (=V/RI_0)$ are obtained by time
averaging on $d\chi/d t'$, for different bias currents $i$. We have
plotted in Fig. (\ref{I-V}) the I-V characteristic for different
values of $i_1$ and $f$. For easier comparison we have juxtaposed
the frames of $f\sim 0$, $f=0.5$ and $f=1.5$. For $i_1\neq 0$ and
$f\neq 0$, as a result of the molecular nanomagnet's spin nutation,
Shapiro-like steps appear at voltages $v=nf$, where $n$ is an
integer number (see Fig. \ref{I-V}-(b),(c)). The width and heights
of these steps are directly controlled by two parameters $\delta
v=f$ and $\delta i\propto i_1$. These Shapiro-like steps disappear
when $\omega$ or $\delta\theta$ is zero, i.e. when the molecular
nanomagnet has only a precession around $z$ axis. Indeed the spin
nutation of the molecular nanomagnet is essential for the emergence
of the Shapiro-like steps on the I-V characteristic. To become more
clear we glance at the time dependence of the phase difference. As a
result of the oscillatory term of the Josephson current, $\chi(t)$
is given by:
\begin{eqnarray}
\chi(t')=\chi_0+v t'+\delta\chi\sin f t'\ ,
\label{phase}
\end{eqnarray}
where $\chi_0$ is a constant, $v$ is the rescaled time-averaged
voltage and $\delta\chi$ is the amplitude of the phase modulation.
Inserting this expression in Eq.(\ref{scaled-RSJ_model}) the time
dependent part of the Josephson current, $i_1\cos f t'\sin\chi$,
can be expanded as
\begin{eqnarray}
i_1 e^{\pm i\chi\pm i f
t'}=i_1 \sum_{n=-\infty}^{+\infty}J_n(\delta\chi)e^{i(\pm\chi_0\pm v
t'\mp n f t')}\ ,\label{Expansion}
\end{eqnarray}
where $J_n(x)$ is an ordinary Bessel function.  The {\it dc}
component of the Josephson current is obtained by time averaging on
the right hand side of Eq. (\ref{Expansion}). The time average is
non-zero at voltages $v=nf$ ($V=n\frac{\hbar\omega}{2e}$) where the
Shapiro-like steps appear on the I-V curve.

According to ferromagnetic-resonance experiments (for example see
Ref.[\onlinecite{Bell08}]), the typical values of the precession
frequencies of a ferromagnetic thin film in the proximity of a
superconductor are in the range of tens of GHz which corresponds to
a magnetic field of about 100 Gauss. Therefore the order of
magnitude of the molecular nanomagnet's precession and nutation
energies, $\hbar\Omega$ and $\hbar\omega$, are about $\sim 10^{-6}
e{\rm V}$ which result in Shapiro-like steps with width of $\sim
1\mu {\rm V}$. The heights of the Shapiro-like steps are
proportional to the nutation amplitude $\delta\theta$, which can be
adjusted by the amplitude and frequencies of the circularly
polarized magnetic fields ($\delta\theta=\gamma h_{xy}/\omega$). In
our results the current and voltage has been rescaled by the $dc$
josephson current $I_0$ and $R I_0$, respectively. The typical
values of $I_0$ and $R$ are in the range of $\sim n {\rm
A}$\cite{Gaudenzi15} and $1$ Ohm which result in the first step
hight to be of the order of $1 n {\rm A}$. The order of magnitude of
the width and heights of Shapiro-like steps are in ranges that are
simply accessible in the experiment, thereby the realization of the
effect presented above is feasible.

Now let us give a discussion on the possibility of the molecular
nanomagnet-superconductor Josephson junction fabrication. In spite
of several transport measurements on molecular nanomagnets connected
to normal leads,\cite{Grose08,Jo06,Heersche06,Zyazin12} the only
practical transport measurements on the molecular nanomagnets
coupled to the superconducting leads has been performed for the
magnetic $C_{60}$ fullerene injected in the electromigrated gold
break junctions.\cite{Kasumov05,Winkelmann09} Superconductivity is
typically induced in these junctions by means of the proximity
effect and molecular nanomagnets are coupled to the gold nanowires.
Very recently it has been demonstrated that electronic transport
through molecular nanomagnets can be extended to superconducting
electrodes by combining gold with molybdenum-rhenium
(MoRe).\cite{Gaudenzi15} This combination induces proximity-effect
superconductivity in the gold to temperatures of at least $4.6 {\rm
K}$ and magnetic fields of $6 \rm{T}$, improving on previously
reported aluminum based superconducting
nanojunctions.\cite{Kasumov05,Winkelmann09}

Recently, it has been shown that a {\it spin} current is generated
in the superconducting leads by spin precession of the molecular
nanomagnet.\cite{Teber10,Holmqvist11,Holmqvist12} The flowing spin
current induces a torque on the nanomagnet and changes its time
evolution. This back-action effect alters the parameters of the
{spin dynamics} ($\Omega$ and $\omega$).\cite{Holmqvist12,Zhu04} It
should be noted that although the back-action affects the motion of
the nanomagnet's spin, it does not disturb the spin nutational
motion. Therefore, we expect that the Shapiro-like steps appear at
the voltages slightly different from $V=n\frac{\hbar\omega}{2e}$.
The interaction of the spin with its environment such as exchange
field and magnetic anisotropy can be considered by introducing a
Gilbert damping constant.\cite{Gatt07} In this case, the spin
dynamics of the molecular nanomagnet is given by the
Landau-Lifshits-Gilbert equation.\cite{Takahashi07} Presence of
the damping will change the amplitude of the nutation, but
not its frequency.\cite{Takahashi07} Thus, the interactions alters
the heights of the steps and not their width, thereby they can not
disturb the observed stepwise structure.

{\it Conclusion}: In conclusion, we have presented a feasible way to
directly detect the spin dynamics of a nutating molecular
nanomagnet. By fabricating a molecular nanomagnet-superconductor
Josephson junction and measuring the voltage difference and
Josephson current through the junction, we are able to obtain the
current-voltage characteristic of the junction. The interplay of the
Josephson current and the spin nutation shows itself as a stepwise
structure on the I-V characteristic, in which the width and heights
of the steps give directly the precession and nutation frequencies
of the molecular nanomagnet's spin. Our results would be of quite
important for experimentalists and provide a direct way to detect
the spin dynamics of a single molecular nanomagnet.

%%%%%%%%%%%%%%%%%%%%%%%%%%%%%%%%%%%%%%%%%%%%%%%%%%%%%%%%%%%%%%%%%%%%%%
%%%%%%%%%%%%%%%%%%%%%%%%%%%%%%%%%%%%%%%%%%%%%%%%%%%%%%%%%%%%%%%%%%%%%%

\section*{References}

\end{document}